\documentclass[twocolumn,showpacs,preprintnumbers,superscriptaddress]{revtex4}
\usepackage{amsmath}
\usepackage{amssymb}
\usepackage{mathrsfs}
\usepackage{graphicx}
\usepackage{bm}
\usepackage{braket}
\usepackage[colorlinks=true,citecolor=blue,linkcolor=blue,urlcolor=blue]{hyperref}

\begin{document}

\title{Exciton states in a circular graphene quantum dot: magnetic field induced intravalley to intervalley transition}

\author{L. L. Li}
\email{longlong.li@uantwerpen.be}
\affiliation{Department of Physics, University of Antwerp,
Groenenborgerlaan 171, B-2020 Antwerpen, Belgium}
\affiliation{Key Laboratory of Materials Physics, Institute of Solid State Physics, Chinese Academy of Sciences, Hefei 230031,
China}
\author{M. Zarenia}
\email{mohammad.zarenia@uantwerpen.be}
\affiliation{Department of Physics, University of Antwerp,
Groenenborgerlaan 171, B-2020 Antwerpen, Belgium}
\author{W. Xu}
\affiliation{Key Laboratory of Materials Physics, Institute of Solid State Physics, Chinese Academy of Sciences, Hefei 230031,
China}
\affiliation{Department of Physics, Yunnan University, Kunming 650091, China}
\author{H. M. Dong}
\affiliation{Department of Physics, China University of Mining and Technology, Xuzhou 221116, China}
\author{F. M. Peeters}
\email{francois.peeters@uantwerpen.be}
\affiliation{Department of Physics, University of Antwerp,
Groenenborgerlaan 171, B-2020 Antwerpen, Belgium}

\date{\today}

\begin{abstract}
The magnetic-field dependence of the energy spectrum, wave function, binding energy and oscillator strength of exciton states confined in a circular graphene quantum dot (CGQD) are obtained within the configuration interaction (CI) method. We predict that: (1) excitonic effects are very significant in the CGQD as a consequence of a combination of geometric confinement, magnetic confinement and reduced screening; (2) two types of excitons (intravalley and intervalley excitons) are present in the CGQD because of the valley degree of freedom in graphene; (3) the intravalley and intervalley exciton states display different magnetic-field dependencies due to the different electron-hole symmetries of the single-particle energy spectra; (4) with increasing magnetic field, the exciton ground state in the CGQD undergoes an intravalley to intervalley transition accompanied by a change of angular momentum; (5) the exciton binding energy does not increase monotonically with the magnetic field due to the competition between geometric and magnetic confinements; and (6) the optical transitions of the intervalley and intravalley excitons can be tuned by the magnetic field and valley-dependent excitonic transitions can be realized in CGQD.

\end{abstract}

\maketitle

\section{Introduction}
Graphene is an atomically thin two-dimensional (2D) crystal made of carbon atoms that are arranged in a honeycomb network. Since its first isolation in 2004 \cite{Novoselov2004} and later experimental demonstration of its excellent transport properties in 2005 \cite{Novoselov2005,Zhang2005}, this 2D atomic crystal
has drawn extensive research attention up to current days. Graphene has a unique electronic structure with zero energy gap and linear energy dispersion, which leads to fascinating physical properties \cite{CastroNeto2009,Abergel2010,DasSarma2011,Kotov2012}
as well as potential device
applications \cite{Schwierz2010,Avouris2010,Bonaccorso2010}. In
recent years, there was considerable interest in quantum confinement
effects in graphene nanostructures. It is expected that they will
modify the physical properties of Dirac fermions in graphene and thus may bring about new quantum phenomena. Due to the Klein tunneling effect, it is impossible to confine carriers in graphene via electrostatic gating \cite{Katsnelson2006,Pereira2006}. However,
lithographic etching of a graphene layer into narrow stripes or small flakes will force carriers into a small area. With current nanofabrication techniques, various graphene nanostructures can be experimentally realized and a number of experimental results have been reported for etched graphene nanostructures \cite{Han2007,Chen2007,Ponomarenko2008,
Stampfer2008,Ritter2009,Guttinger2009}. An alternative route is the chemical assembly of carbon atoms into small structures such as short nanoribbons and dot-like structures with well-defined edge structure \cite{CaiJ2010,LiuR2011}. Among these nanostructures,
graphene quantum dots (GQDs) are of particular interest because they
exhibit excellent electronic and optical properties which can be tuned by changing their lateral size, geometric shape,
boundary type, sublattice symmetry and the number of graphene
layers \cite{Yamamoto2006,Heiskanen2008,Zhang2008,
Akola2008,Tang2008,Peres2009,Libisch2009,Libisch2010,
Recher2010,Zarenia2011,RozhkovA2011,Guttinger2012,Guclu2013a,
Guclu2013,Yamijala2015,Hawrylak2016}. Moreover, due to their excellent and tunable electronic and optical properties, GQDs hold promising applications in advanced electronics and optoelectronics. A comprehensive review of the current status of GQDs can be found in Ref. \cite{Guclu2014}.

Many-body effects such as excitonic effects induced by electron-hole
interactions are expected to be interesting and important in
graphene due to its 2D character and reduced screening. A number of
theoretical \cite{Yang2009,Peres2010,Trevisanutto2010} and
experimental \cite{Mak2011,Chae2011,Yadav2015} studies have revealed
that remarkable excitonic effects are indeed observed in the optical
absorption spectrum of graphene. This indicates that one has to go
beyond the single-particle picture in order to accurately describe
the optical properties of graphene. Despite the considerable number of studies on excitonic effects in graphene \cite{Yang2009,Peres2010,Trevisanutto2010,
Mak2011,Chae2011,Yadav2015}, less attention has been paid on the exciton problem in GQDs. Compared with bulk graphene, GQDs have finite energy gaps and exhibit carrier confinement, which can lead to enhanced electron-hole interaction and thus result in stronger
excitonic effects. Up to date, the exciton problem in GQDs has been
investigated in only a few theoretical studies \cite{Guclu2010,Ozfidan2015,Li2015}. However, in
these studies the effect of an external magnetic field on the exciton states has not been explored. In the present work, we theoretically investigate the exciton states in a model circular GQD (CGQD) with the infinite-mass boundary condition in the presence of a perpendicular magnetic field. The infinite-mass boundary condition states that the outward current at the dot edge is zero \cite{Berry1987}, which can be realized by applying an infinite staggered potential outside the dot. Consequently, the particular edge which may play an important role in realist GQDs is no longer important in the present circular model. The accurate treatment of edges requires an analysis based on the tight-binding model or first-principle calculations. Although the circular model is perhaps the simplest model, it captures the main qualitative physics in graphene dots and can be both analytically and numerically solved. Moreover, it can provide a good starting point to study both the single-particle and many-body properties of QDs in graphene.

In this work, we show that apart from the intrinsic geometrical confinement, the extrinsic magnetic confinement has also a significant influence on the exciton states in the CGQD. To calculate these many-body states in the considered system, the following two steps are carried out: First, the single-particle states of electrons and holes are calculated by solving the Dirac equation with infinite-mass boundary condition; Second, using these single-particle states, the configuration-interaction (CI) method \cite{Bryant1987} is employed to calculate the exciton states induced by the electron-hole interaction. Within the CI method, the exciton wave function in the CGQD is expanded as a linear combination of products of the electron and hole single-particle wave functions which may reside in one of the two valleys of graphene. The results for the magnetic field dependence of the exciton states is presented and discussed, and some interesting features are observed. We show that due to the valley degree of freedom in graphene, the exciton states in the CGQD are more complicated than those in a conventional semiconductor quantum dot (CSQD), because in the CGQD the electrons and holes have the possibility to be in the same valley or in different valleys.

This paper is organized as follows. In Section II, we present the
theoretical model and calculation method for the exciton states in a
CGQD in the presence of a magnetic field. In Section III, the
numerical results on the magneto-exciton states are presented and discussed. Finally, our concluding
remarks are given in Section IV.

\section{Model and Theory}
Our theoretical approach is divided into two parts: in the first
part we employ the Dirac equation to calculate the single-particle
energies and wave functions of confined electrons and holes in both
valleys, and in the second part we use the configuration-interaction (CI) method to calculate the exciton states by including the electron-hole interaction and expanding the exciton wave function in terms of the electron and hole single-particle wave functions obtained in the first part.

\begin{figure}[t]
\begin{center}
\includegraphics[width=0.49\textwidth]{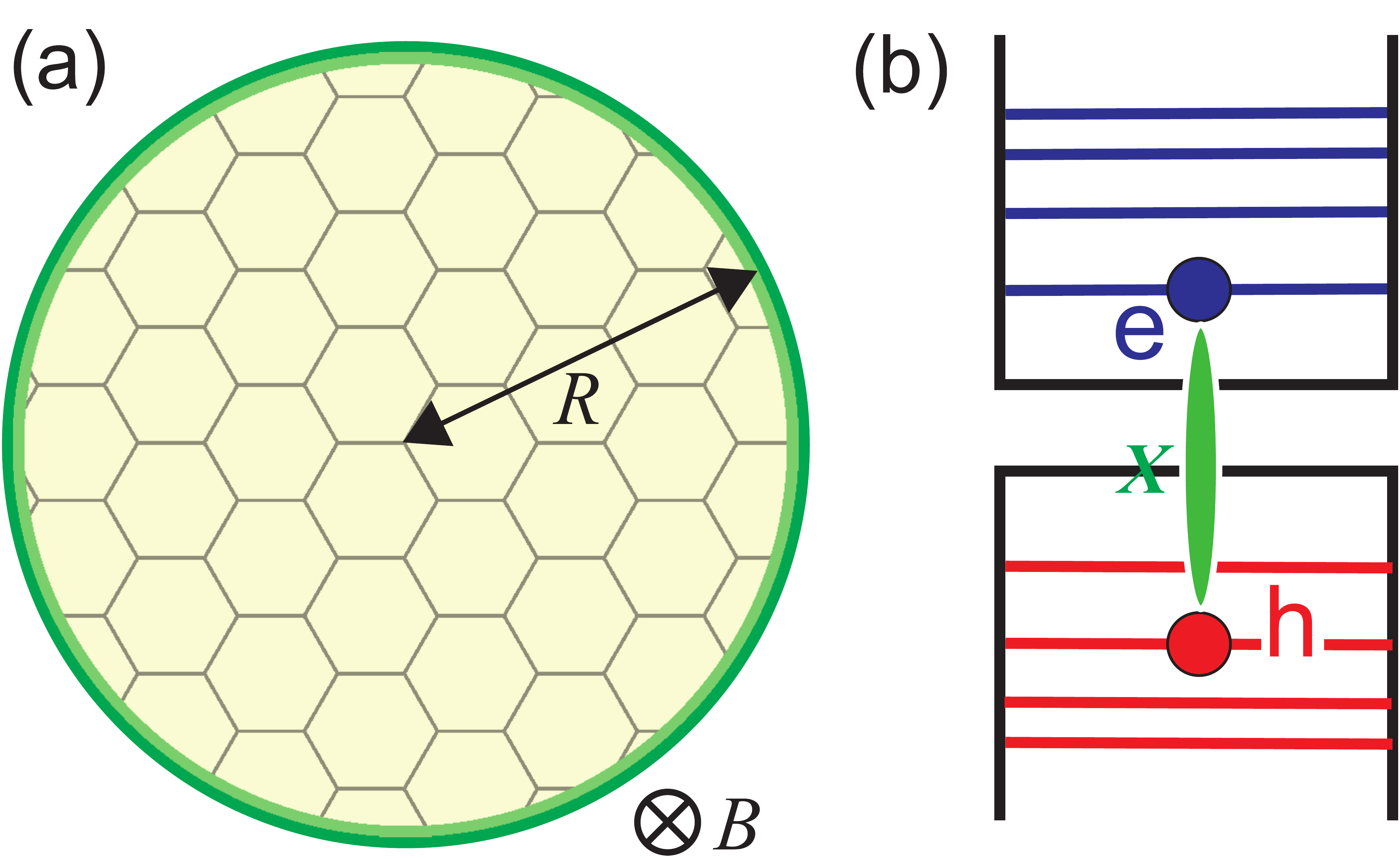}
\caption{Model system considered in the present work: (a) a circular graphene quantum dot (CGQD) of radius $R$ in the presence of a perpendicular magnetic field $B$ and (b) an exciton ($X$) formed in the CGQD by a conduction electron ($e$) and a valence hole ($h$) via the attractive Coulomb interaction.} \label{Fig1}
\end{center}
\end{figure}

\subsection{Single-particle states of confined electrons and holes}
We consider a CGQD of radius $R$ in the presence of a perpendicular
magnetic field $B$, as illustrated in Fig. 1. In order to obtain the electron and hole single-particle states in the considered system, we use the Dirac Hamiltonian describing the low-energy dynamics of electrons and holes in graphene. In the valley-isotropic form, this Hamiltonian is given by \cite{Schnez2008,Grujic2011}
\begin{equation}\label{e1}
H=v_F(\textbf{p}+e\textbf{A})\cdot\boldsymbol{\sigma}+\tau
V(\textbf{r})\sigma_z,
\end{equation}
where $v_F$ is the Fermi velocity of graphene, $\textbf{p}=(p_x,p_y)$ is the in-plane momentum operator with $p_x=-i\hbar\partial/\partial x$, $\textbf{r}=(x,y)$ is the in-plane position vector, $\textbf{A}=(-By/2,Bx/2)$ is the magnetic vector potential in the symmetric gauge, $\boldsymbol{\sigma}=(\sigma_x,\sigma_y,\sigma_z)$ is the Pauli matrix vector, $\tau$ is the valley index of graphene with $\tau=+1$
($-1$) denoting the $K$ ($K'$) valley, and $V(\textbf{r})$ is the
mass-related potential. We assume the charge carriers (electrons and holes) are confined in the CGQD, which can be modeled by a zero (infinite) potential inside (outside) the CGQD \cite{Schnez2008,Grujic2011}, i.e., $V(\textbf{r})=0$ for $|\textbf{r}|<R$ and $V(\textbf{r})=\infty$ for $|\textbf{r}|\geq R$. Note that the confinement potential $V(\textbf{r})$ appears in the Dirac Hamiltonian (1) via the Pauli matrix $\sigma_z$, so it adds only the diagonal terms in this $2\times 2$ Hamiltonian. When both the inequivalent $K$ and $K'$ valleys are included, the original $2\times 2$ Hamiltonian will become a $4\times 4$ one, but the confinement potential $V(\textbf{r})$ still appears in the diagonal terms of this new Hamiltonian, and there are no off-diagonal terms in the valley basis. Therefore, the $K$ and $K'$ valleys remain decoupled in the presence of confinement. Because the considered system has circular symmetry, it is convenient to adopt cylindrical coordinates, i.e., $\textbf{r}=(r,\phi)$ with $r$ and $\phi$ being the radial coordinate and azimuthal angle in the 2D plane, respectively.

The single-particle states of confined electrons and holes can be obtained by solving the Dirac equation $H\psi=E\psi$, where $E$ and $\psi$ are the single-particle energy and wave function, respectively. To solve this equation, we introduce dimensionless variables $\rho=r/R$, $\beta=R^2/(2l_B^2)$ and $\varepsilon=ER/(\hbar v_F)$, where $l_B=\sqrt{\hbar/(eB)}$ is the magnetic length with $e$ and $\hbar$ being the elementary charge and the reduced Planck constant, respectively. With these dimensionless variables, the Dirac equation $H\psi=E\psi$ in cylindrical coordinates can be written as
\begin{equation}\label{e2}
\left[\begin{array}{cc}
0 & \pi_{-} \\
\pi_{+} & 0 \\
\end{array}\right]
\left[\begin{array}{cc}
\psi_1(\rho,\phi) \\
\psi_2(\rho,\phi)\end{array}\right]=\varepsilon\left[\begin{array}{cc}
\psi_1(\rho,\phi) \\
\psi_2(\rho,\phi)\end{array}\right],
\end{equation}
where $\pi_{\pm}=-i e^{\pm i\phi}[\partial/\partial
\rho\pm(i/\rho)\partial/\partial \phi\mp \beta\rho]$ and
$\psi_j(\rho,\phi)$ ($j=1,2$) are the two components of the
wave function $\psi(\rho,\phi)$. Due to the circular symmetry of our problem, the two components of the wave function $\psi(\rho,\phi)$ can be written as
\begin{equation}\label{e3}
\left[\begin{array}{cc}\psi_1(\rho,\phi) \\
\psi_2(\rho,\phi)\end{array}\right]
=\frac{1}{\sqrt{2\pi \mathcal{N}}}e^{im\phi}\left[\begin{array}{cc}u(\rho) \\
ie^{i\phi}v(\rho)\end{array}\right],
\end{equation}
where $m$ is the angular quantum number which takes integer values, and $\mathcal{N}$ is the normalization factor determined by the normalization condition for the wave function, i.e., $\mathcal{N}=R^2\int_0^1 (|u(\rho)|^2+|v(\rho)|^2)\rho d\rho$. The two components of the wave function correspond to different sublattice contributions, i.e., $u(\rho)$ corresponds to the contribution from sublattice $A$ and $u(\rho)$ to that from sublattice $B$. Inserting the two-component wave function (3) into the Dirac equation (2), we obtain the following set of coupled ordinary differential equations:
\begin{equation}\label{e4}
\left[\begin{array}{cc}
0 & f(\beta,m,\rho) \\
g(\beta,m,\rho) & 0 \\
\end{array}\right]
\left[\begin{array}{cc} u(\rho) \\ v(\rho)\end{array}\right]
=\varepsilon\left[\begin{array}{cc}u(\rho) \\
v(\rho)\end{array}\right],
\end{equation}
where $f(\beta,m,\rho)=\partial/\partial\rho+(m+1)/\rho+\beta\rho$ and $g(\beta,m,\rho)=-\partial/\partial\rho+m/\rho+\beta\rho$.
To solve these equations, we still need some boundary conditions. The mass-related potential $V(\textbf{r})$ in the CGQD leads to the infinite-mass boundary condition \cite{Berry1987}, which requires that the outward current at the dot edge is zero and yields the simple condition $\psi_2(\rho=1,\phi)/\psi_1(\rho=1,\phi)=i\tau
e^{i\phi} $ or $v(\rho=1)/u(\rho=1)=\tau $ for circular
confinement \cite{Berry1987,Schnez2008,Grujic2011}. It should be noted that this boundary condition for the CGQD is quite different from that for the CSQD, which requires the wave function (not the current) to vanish at the dot boundary. The single-particle states of confined electrons and holes in the CGQD in the presence of perpendicular magnetic field are obtained by numerically solving the coupled differential equations (4) with the infinite-mass boundary condition using the finite element method \cite{Zarenia2011}. The obtained single-particle states are characterized by the set of quantum numbers $(\tau,m,n)$, where $\tau$ is the valley index, $m$ is the angular quantum number and $n$ is the principal (or radial) quantum number. For large magnetic fields, $n$ can be identified as the Landau level index.

\subsection{Exciton states induced by electron-hole interactions}

After obtaining the single-particle states of electrons and holes,
we now consider the exciton states induced by electron-hole
interactions in a CGQD in the presence of a magnetic field. The
exciton Hamiltonian ($H_X$) for the model system is given by
\begin{equation}\label{e5}
H_{X}=H_e+H_h+V_{eh},
\end{equation}
where $H_e$ ($H_h$) is the single-particle Hamiltonian for the
electron (hole), $V_{eh}=-e^2 /(4\pi\kappa|\textbf{r}_e-\textbf{r}_h|)$
is the Coulomb interaction between the electron and the hole,
$\textbf{r}_e$ ($\textbf{r}_h$) is the electron (hole) coordinates, and $\kappa$ is the effective dielectric constant of graphene. Note that $V_{eh}$ is the unscreened (bare) electron-hole Coulomb interaction. It has been shown \cite{Berkelbach2013} that in atomically thin 2D materials such as monolayer MoS$_2$, the electron-hole Coulomb interaction can be taken as of the Keldysh type in which the nonlocal screening effect is properly taken into account. Here we limit ourselves to the unscreened Coulomb interaction between an electron and a hole and we do not expect any qualitative changes if the Coulomb potential is modified.

The exciton states can be obtained by solving the two-particle Schr\"odinger equation $H_{X}\Psi_{X}(\textbf{r}_e,\textbf{r}_h)=
E_{X}\Psi_{X}(\textbf{r}_e,\textbf{r}_h)$, where $E_{X}$ and $\Psi_{X}$ are the exciton energy and wave function, respectively. In the present work, we employ the CI method \cite{Bryant1987} to calculate the exciton states. In this method, the exciton wave function is expanded as a linear combination of direct products of the electron and hole wave functions. To proceed, we define two important quantities for the exciton wave function: the total valley index $T=\tau_e+\tau_h$ and the total angular momentum $M=m_e+m_h$ for the electron-hole pair, where $\tau_j$ and $m_j$ ($j=e, h$) are the valley index and the angular momentum for the single-particle state, respectively. With this definition, we may expand the exciton wave function with fixed $T$ and $M$ as
\begin{equation}\label{e6}
\Psi_{X}(\textbf{r}_e,\textbf{r}_h)=\sum_{\lambda_e\lambda_h}
A_{\lambda_e\lambda_h}\psi_{\lambda_e}(\textbf{r}_e)\psi_{\lambda_h}(\textbf{r}_h),
\end{equation}
where $A_{\lambda_e\lambda_h}$ is the expansion coefficient, and the
subscripts $\lambda_e=(\tau_e,m_e,n_e)$ and $\lambda_h=(\tau_h,m_h,n_h)$ are the quantum number sets for the
electron and hole single-particle states, respectively.
Electron-hole pairs in the summation of Eq. (\ref{e6}) are limited to those satisfying $\tau_e+\tau_h=T$ and $m_e+m_h=M$. With this expansion of the exciton wave function, the exciton Schr\"odinger equation now reads
\begin{equation}\label{e7}
(E_{\lambda_e'}+E_{\lambda_h'}-E_X)A_{\lambda_e'\lambda_h'}+\sum_{\lambda_e\lambda_h}
V_{\lambda_e\lambda_h}^{\lambda_e'\lambda_h'}A_{\lambda_e\lambda_h}=0,
\end{equation}
where $E_{\lambda_e'}$ ($E_{\lambda_h'}$) is the single-particle
energy of the electron (hole) state, and the Coulomb matrix element
$V_{\lambda_e\lambda_h}^{\lambda_e'\lambda_h'}$ is given by
\begin{equation}\label{e8}
V_{\lambda_e\lambda_h}^{\lambda_e'\lambda_h'}=\int\int
\psi_{\lambda_e'}^{\dag}(\textbf{r}_e)\psi_{\lambda_h'}^{\dag}(\textbf{r}_h)
V_{eh}\psi_{\lambda_e}(\textbf{r}_e)\psi_{\lambda_h}(\textbf{r}_h)
d\textbf{r}_ed\textbf{r}_h.
\end{equation}
In the derivation of Eq. (7), the orthogonality of the
single-particle wave function has been used, i.e.,
$\braket{\lambda_j'|\lambda_j}=\delta_{\lambda_j',\lambda_j}$ ($j=e, h$) with $\delta$ being the Kronecker delta. Because the considered system has circular symmetry, the total exciton angular momentum $M$ is a conserved quantity, and thus the Coulomb matrix element given by Eq. (8) is nonzero only when $M'=M$ ($M'=m_e'+m_h'$ and $M=m_e+m_h$). After calculating all nonzero Coulomb matrix elements, the full exciton Hamiltonian matrix is then diagonalized to obtain the eigenvalues (corresponding to the exciton energy levels) and eigenvectors (corresponding to the expansion coefficients for the exciton wave functions). In the numerical diagonalization, the basis states used in the CI method are chosen such that they are the lowest single-particle states of electrons and holes and the number of these states is chosen sufficiently large to guarantee convergence of the lowest exciton energies. The singularity occurring in the Coulomb matrix element can be removed by using an alternative expression in terms of the Legendre function of the second kind of half-integer degree \cite{Cohl2001}.

Given the exciton energy spectrum and corresponding wave function, different physical properties of the exciton can be evaluated in
principle. Here, we present the binding energy, effective radius,
and oscillator strength of the exciton. These physical quantities
are very helpful in understanding the excitonic properties of the
material system. The exciton binding energy $E_B$, effective radius $R_X$, and oscillator strength $F_X$ are given by \cite{TadicM2011}
\begin{equation}\label{e9}
E_B=\bra{\Psi_X}H_e+H_h\ket{\Psi_X}-E_{X},
\end{equation}
\begin{equation}\label{e10}
R_X=\sqrt{\bra{\Psi_X}|\textbf{r}_e-\textbf{r}_h|^2\ket{\Psi_X}},
\end{equation}
and
\begin{equation}\label{e11}
F_X=\Big{|}\int\int\Psi_X(\textbf{r}_e,\textbf{r}_h)
\delta(\textbf{r}_e-\textbf{r}_h)d\textbf{r}_ed\textbf{r}_h\Big{|}^2,
\end{equation}
respectively. As can be derived from the expression of $F_X$, the
exciton states with total angular momentum $M=0,-1,-2$ contribute to the optical transitions, i.e., they are optically active (or bright) states. This result for the CGQD is different from that for the CSQD where only exciton states with $M=0$ are optically
bright \cite{Shi2002}. In the present work, we will limit
ourselves to the optically bright exciton states since they can be
experimentally observed in photoluminescence spectra.

\begin{figure}[htbp]
\begin{center}
\includegraphics[width=0.48\textwidth]{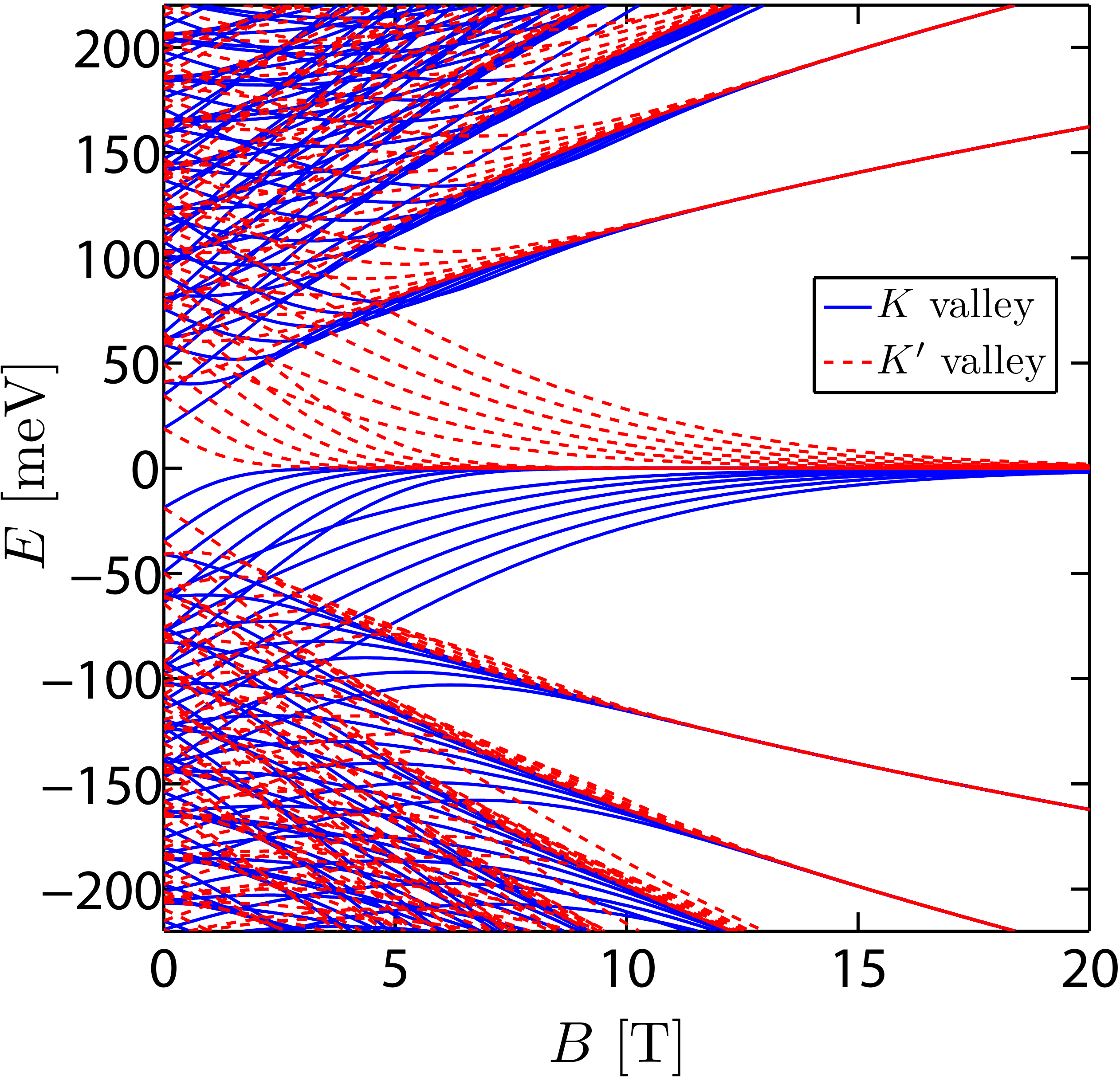}
\caption{Single-particle energy spectrum in a CGQD with $R=50$ nm in
the presence of a perpendicular magnetic field. Only six lowest electron and hole states are shown for the angular quantum number $-6\leq m\leq 6$. The blue solid and red dashed curves denote the results for the $K$ and $K'$ valleys, respectively, as indicated.} \label{Fig2}
\end{center}
\end{figure}

In addition to the binding energy, effective radius and oscillator
strength, we also present the electron-hole pair density and
the conditional probability density, which are given by \cite{YannouleasC2000}
\begin{equation}\label{e12}
n(r)=\sum_{j=e,
h}\bra{\Psi_X}\delta(\textbf{r}-\textbf{r}_j)\ket{\Psi_X},
\end{equation}
and
\begin{equation}\label{e13}
P(\textbf{r}_h|\textbf{r}_e=\textbf{r}_0)=\frac{\big{|}\Psi_{X}(\textbf{r}_e=
\textbf{r}_0,\textbf{r}_h)\big{|}^2}{\int\big{|}\Psi_{X}(\textbf{r}_e=
\textbf{r}_0,\textbf{r}_h)\big{|}^2d\textbf{r}_h},
\end{equation}
respectively. These two quantities are very useful in characterizing
the spatial distribution of the exciton state. According to their
definitions, $n(r)$ gives the electron-hole pair density at a
radial distance $r=|\textbf{r}|$, while $P(\textbf{r}_h|\textbf{r}_e=\textbf{r}_0)$ gives the
probability to find the hole at $\textbf{r}_h $ under the condition
that the electron is pinned at $\textbf{r}_e$.

\section{Results and discussions}

\begin{figure*}[htbp]
\begin{center}
\includegraphics[width=0.75\textwidth]{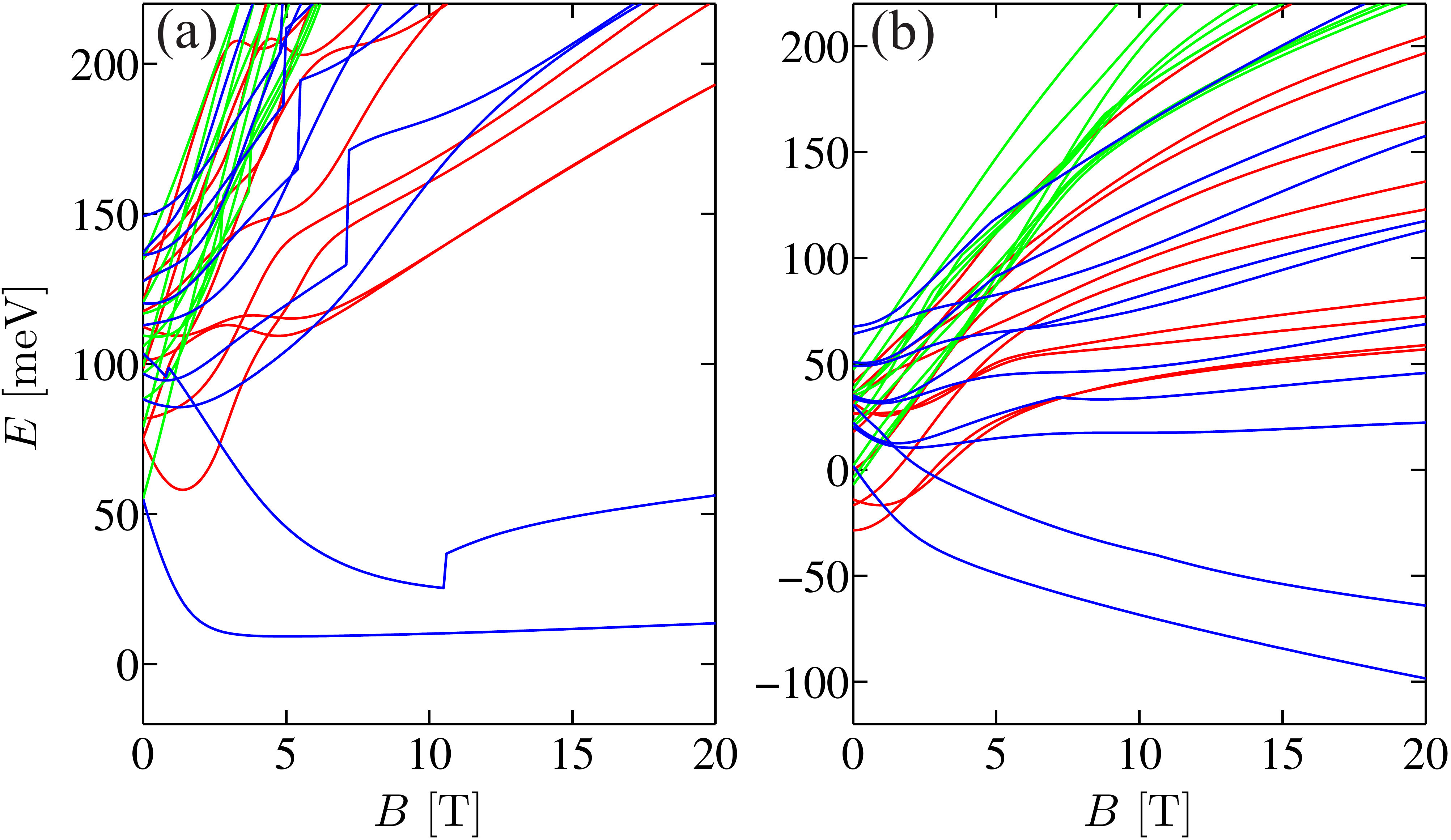}
\caption{Magnetic energy spectra of non-interacting states (a) and exciton states (b) for the same CGQD as in Fig. 2. Energy levels with the same total valley index $T$ are plotted in the same type of
curve: the red curves for $T=\pm2$ with $\tau_e=\tau_h=\pm1$, the green curves for $T=0$ with $\tau_e=-\tau_h=1$, and the blue curves also for $T=0$ but with $\tau_e=-\tau_h=-1$. Note that the energy levels for $T=\pm2$ are degenerate due to the same electron-hole symmetry and thus are plotted with the same red color.} \label{Fig3}
\end{center}
\end{figure*}

In this part, we will present and discuss our numerical results
for the single-particle states of confined electrons and holes and
for the exciton states induced by electron-hole interactions in a
CGQD in the presence of a perpendicular magnetic field. For the numerical calculation of exciton states in the present work, we choose the angular quantum number $m_e=m_h=0,\pm1,\pm2,...,\pm7$ and the radial quantum number $n_e=n_h=1, 2, ..., 7$ in Eq. (7) for both the $K$ ($\tau=1$) and $K'$ ($\tau=-1$) valleys, which gives an accuracy for the exciton ground-state energy close to $10^{-3}$ meV.
Furthermore, we take $\kappa=2.5$ for the effective dielectric
constant of graphene \cite{Hwang2007,Abergel2008}. Such a small
dielectric constant leads to reduced screening for the Coulomb
interaction. This is why many-body effects such as electron-electron
and electron-hole interactions are expected to be significant in
graphene.

We first give a brief analysis of the single-particle states in the
CGQD in the presence of a magnetic field which have been analyzed in
detail in a previous work \cite{Grujic2011}. In Fig. 2, we show the
single-particle magnetic energy spectrum in a CGQD with radius
$R=50$ nm. From this figure, we can see the following interesting
features: (1) an energy gap between the electron states in the $K'$ valley and the hole states in the $K$ valley opens at low magnetic fields due to the quantum confinement effect and this gap tends to close as the magnetic field increases; (2) at high magnetic fields, the electron and hole energy levels in both $K$ and $K'$ valleys approach the Landau levels (LLs) of bulk graphene \cite{Grujic2011}. The reason is that the magnetic confinement becomes stronger than the geometric confinement with increasing magnetic field. (3) The low-lying electron LLs (high-lying hole LLs) in the $K$ ($K'$) valley converge to the zero-energy states as the magnetic field increases, which is not seen in the CSQD \cite{Peeters1996} where the energy gap increases with the field. These zero-energy states emerge due to the closure of the energy gap at high magnetic fields. (4) There is intervalley electron-hole symmetry (i.e., $|E_e(\pm\tau,m,n)|=|E_h(\mp\tau,m,n)|$) but no intravalley one
(i.e., $|E_e(\tau,m,n)|\neq|E_h(\tau,m,n)|$) in the CGQD. (5) In the absence of the magnetic field, the energy levels corresponding to the $K$ and $K'$ valleys are degenerate because such two valleys are related to one another by the time-reversal symmetry. This degeneracy is lifted for nonzero magnetic field because the application of this field breaks the time-reversal symmetry.

\begin{figure}[htbp]
\begin{center}
\includegraphics[width=0.48\textwidth]{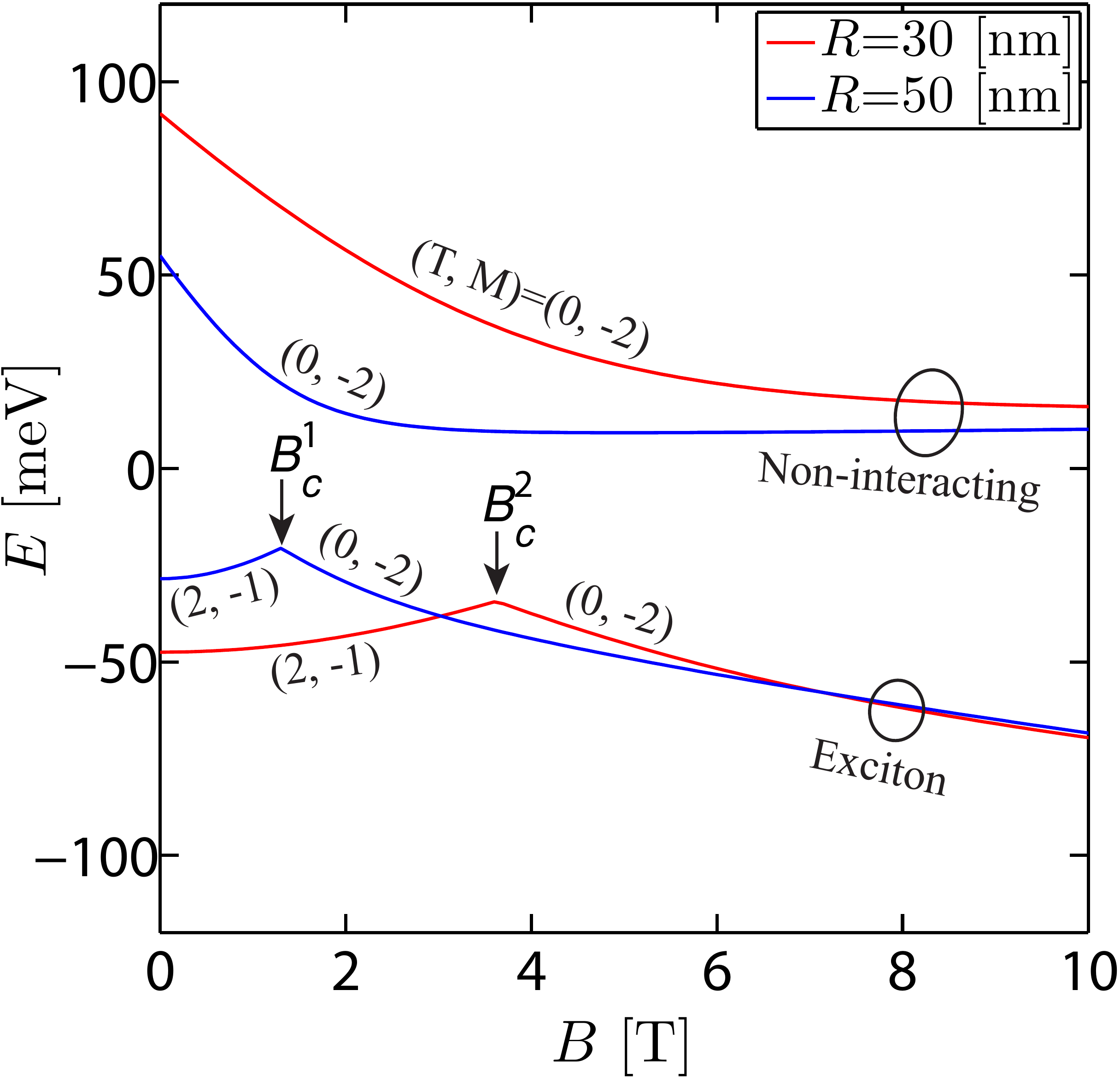}
\caption{Magnetic-field dependencies of non-interacting ground-state
energy and exciton ground-state energy for two dot radii $R$'s as indicated. Here, $(T, M)$ are a pair of quantum numbers: the total valley index and the total angular momentum. The black arrows indicate the critical magnetic fields $B_c^1=1.3$ T for $R=50$ nm and $B_c^2=3.6$ T for $R=30$ nm. At these magnetic fields, the exciton ground state in the CGQD undergoes an intravalley to intervalley transition accompanied by a change of angular momentum, i.e., a combined valley and angular momentum transition.} \label{Fig4}
\end{center}
\end{figure}

Now we turn to the results for the exciton states in the
CGQD in the presence of a magnetic field. To proceed, we first
give some basic physical pictures for exciton formation in the
considered system. In order to make our statements more clearly, we
have to look at the single-particle energy spectrum shown in Fig. 2 again. As can be seen, there can be two types of excitons present in the CGQD: (1) the intravalley exciton formed by an electron and a hole in the same $K$ (or $K'$) valley, and (2) the intervalley exciton formed by an electron in the $K$ (or $K'$) valley and a hole in the $K'$ (or $K$) valley. From the different electron-hole symmetries exhibited in the single-particle energy spectra, we may expect that the intravalley and intervalley exciton states should display different magnetic-field dependencies.

In Fig. 3, we show the magnetic energy spectra of (a) non-interacting states and (b) exciton states for the same CGQD as in Fig. 2. Here, the energy levels with the same total valley index $T$ ($T=\tau_e+\tau_h$) are plotted with the same type of curve: the red solid curves for $T=\pm2$ with $\tau_e=\tau_h=\pm1$, the green solid curves for $T=0$ with $\tau_e=-\tau_h=1$, and the blue solid curves also for $T=0$ but with $\tau_e=-\tau_h=-1$. It's clear that the excitonic effect induced by the electron-hole Coulomb interaction is very significant. We find that for $T=\pm2$ the energy levels are degenerate due to the same electron-hole symmetry for such two valley indices (see Fig. 2). And due to such electron-hole symmetry, the energy spectra for $T=0, \pm2$ exhibit very different magnetic-field dependencies. It should be noted that some energy levels of the non-interacting states [see the blue curves in Fig. 3(a)] exhibit a discontinuous behavior. This is because only diagonal matrix elements of electron and hole angular momenta are involved in the calculation of non-interacting states and thus the total angular momentum $M$ ($M=m_e+m_h$) can be discontinuous as a function of the magnetic field. However, for exciton states, all diagonal and non-diagonal matrix elements are included in the calculation and thus the corresponding energy levels are continuous although some can exhibit a discontinuous derivative.

\begin{figure}[htbp]
\begin{center}
\includegraphics[width=0.48\textwidth]{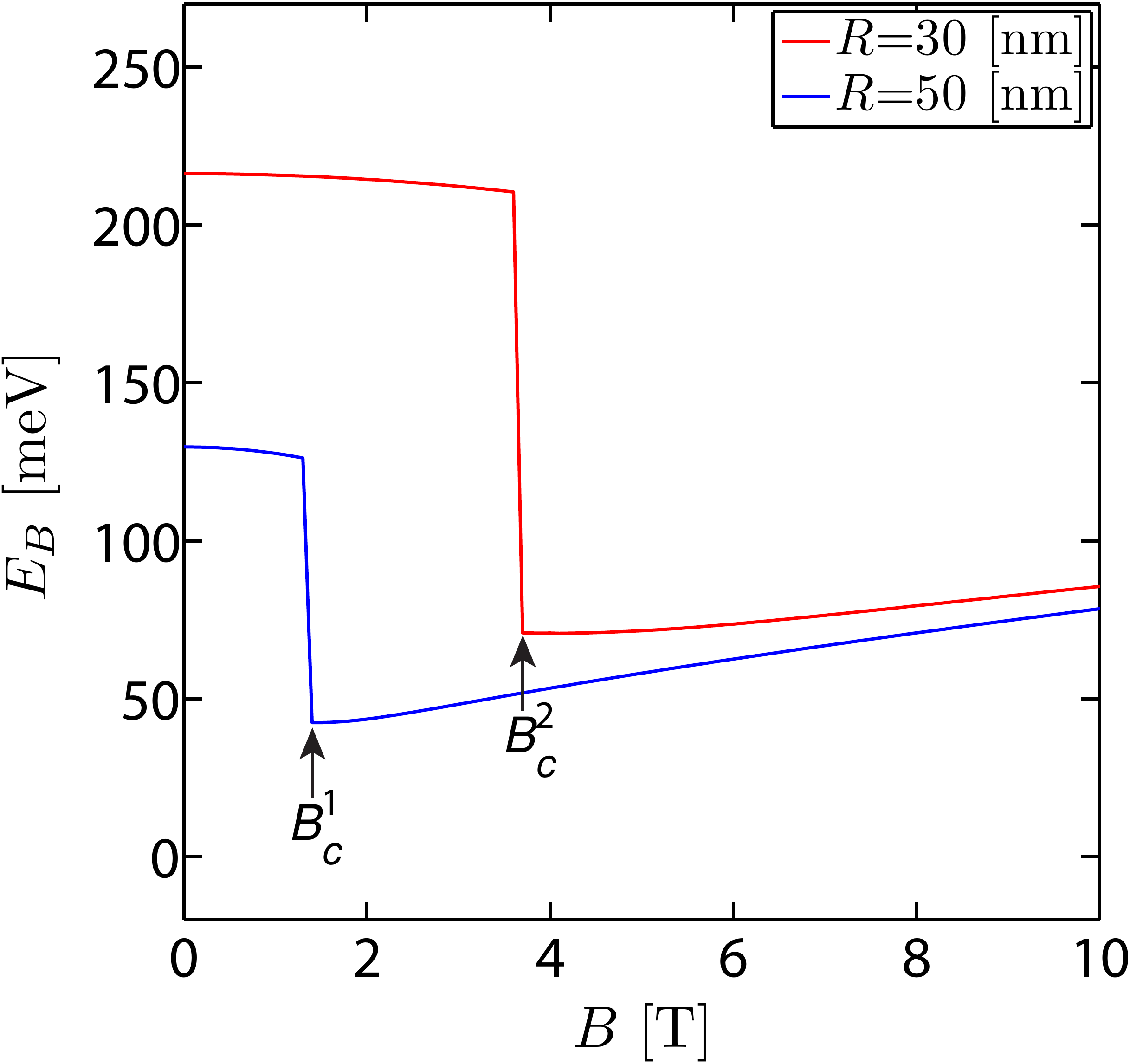}
\caption{Magnetic-field dependence of the binding energy $E_B$ of the exciton ground state for the same CGQDs as in Fig. 4. The black arrows indicate the critical magnetic fields $B_c^1=1.3$ T for $R=50$ nm and $B_c^2=3.6$ T for $R=30$ nm. At these magnetic fields, $E_B$ changes abruptly.} \label{Fig5}
\end{center}
\end{figure}

\begin{figure*}[htbp]
\begin{center}
\includegraphics[width=\textwidth]{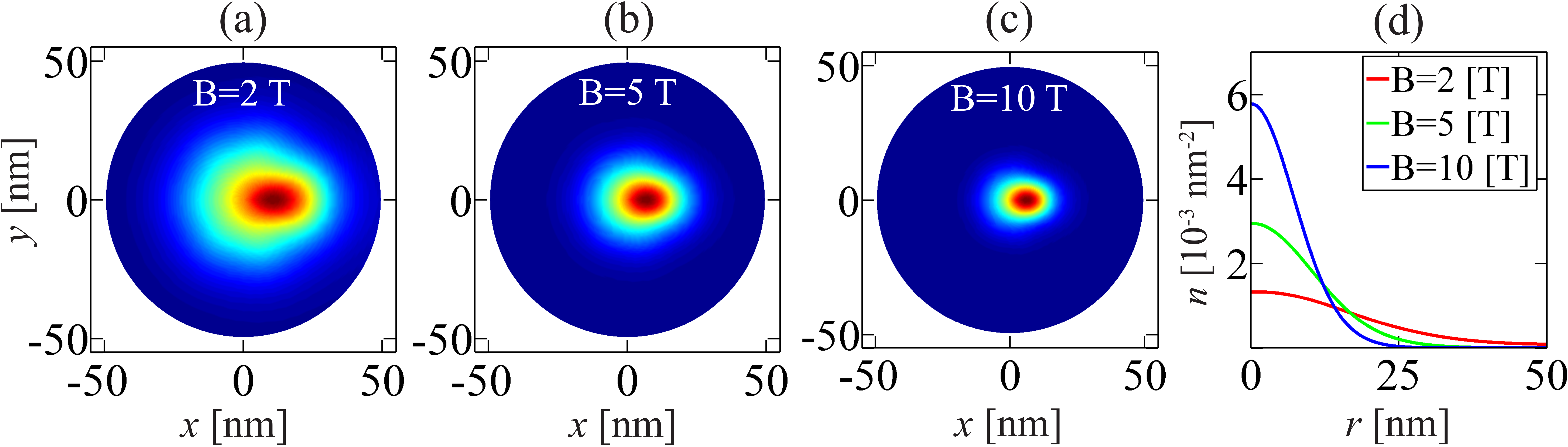}
\caption{Conditional probability densities (CPDs) and electron-hole pair densities (EHPDs) of the exction ground state for the same CGQD as Fig. 2 for different magnetic fields: (a) the CPD at $B=2$ T; (b) the CPD at $B=5$ T; (c) the CPD at $B=10$ T with fixed electron position at $\textbf{r}=(R/2, 0)$; and (d) the EHPDs for these magnetic fields.} \label{Fig6}
\end{center}
\end{figure*}

In Fig. 4, we show the magnetic-field dependence of the exciton
ground-state energy for two dot radii $R$'s as indicated. For comparative purposes, we also plot the non-interacting ground-state energy in the figure. Because the magnetic energy levels for the total valley indices $T=\pm2$ are degenerate (see Fig. 3), we only show the results for $T=2$ in Fig. 4. As can be seen, for both $R=50$ nm and $R=30$ nm, with increasing magnetic field, the exciton ground state in the CGQD undergoes an intravalley to intervalley transition accompanied by a change of angular momentum, i.e., $T=2$ and $M=-1$ becomes $T=0$ and $M=-2$ as the magnetic field increases. For the larger radius ($R=50$ nm), there is a critical value of the magnetic field $B_c^1=1.3$ T at which such a transition occurs, as indicated by the black arrow in the figure. Such a combined transition of valley and angular momentum does not occur in the CSQD where $T$ is not present and $M$ remains unchanged \cite{Janssens2001}, i.e., the exciton state is always a singlet state. But for the smaller radius ($R=30$ nm), the critical magnetic field decreases from $B_c^1=1.3$ T to $B_c^2=3.6$ T. Since the critical magnetic field is determined by the crossing of two exciton energy levels with different valley indices $T$'s and angular momenta $M$'s, the larger energy difference between such two levels corresponds to the larger critical value of the magnetic field. Hence, the critical magnetic field increases as the dot radius decreases. In addition, we find no $T$ and $M$ transitions for the non-interacting ground state for both dot radii ($T=0$ and $M=-2$ are kept when varying the magnetic field). Based on the above statements, we conclude that the intravalley to intervalley transition of the exciton ground state in the CGQD is induced by the electron-hole Coulomb interaction. We also note that at $B=0$, the exciton energy in the CGQD is lower for $R=30$ nm than for $R=50$ nm, which is contrary to most CSQD, where the confinement energy prevails over the Coulomb interaction and would make the exciton energy for $R=30$ nm higher. This difference can be explained as follows. Considering the quadratic low-energy dispersion $E\sim k^2$ in semiconductors, the confinement energy in CSQDs exhibits $1/R^2$ dependence on the dot size $R$ (assuming $k \sim 1/R$), which may prevail over the Coulomb energy ($\sim 1/R$) and would make smaller dot sizes higher in energy. In contrast, the low-energy dispersion is linear $E \sim k$ in graphene, leading to $1/R$ dependence of the confinement energy in a CGQD, which is comparable to the Coulomb energy. Therefore, in the presence of the electron-hole Coulomb interaction, a smaller size of a CGQD may have lower exciton states compared to the one with a larger size.

In Fig. 5, we show the magnetic-field dependence of the binding
energy $E_B$ of the exciton ground state for the same CGQDs as Fig. 4. The smaller dot radius corresponds to the larger exciton binding energy as expected. The binding energy $E_B$ changes abruptly at the critical magnetic fields $B_c^1=1.3$ T and $B_c^2=3.6$ T for the dot radii $R_1=30$ nm and $R_2=50$ nm, respectively. This is a consequence of the intravalley to intervalley transition of the exciton ground state (see Fig. 4). We find that when the magnetic length $l_B=\sqrt{\hbar/eB}$ is comparable to the dot radius $R$, the binding energy $E_B$ changes abruptly at the critical magnetic field $B_c$. The $E_B-B$ relation exhibits different behaviors in the different regions defined by critical magnetic fields. For larger (small) radius $R=50$ nm ($R=30$ nm), $E_B$ decreases slightly with increasing $B$ for $B<B_c^1$ ($B<B_c^2$) and increases markedly with increasing $B$ for $B>B_c^1$ ($B>B_c^2$). At lower magnetic fields, $B<B_c^1$ ($B<B_c^2$) for larger (smaller) $R$, the peculiar $E_B-B$ relation (i.e., $E_B$ decreases slightly with increasing $B$) is mainly caused by the competing effects of geometric and magnetic confinements. However, at higher magnetic fields, $B>B_c^1$ ($B>B_c^2$) for larger (smaller) $R$, $E_B$ increases monotonically with $B$. This is not surprising, because by applying higher magnetic fields the electrons and holes are more confined due to the strong magnetic confinement, they are closer to each other and thus are more tightly bound, which leads to an increase of the exciton binding energy. To see this more intuitively, we plot in Fig. 6 the conditional probability densities (CPDs) and the electron-hole pair densities (EHPDs) of the exciton ground state in the CGQD with radius $R=50$ nm for different magnetic fields as indicated. The expressions for the CPD and EHPD are given, respectively, by Eqs. (12) and (13) in Section II. In this figure, we can see that with increasing magnetic field, the electrons and holes are more confined in the CGQD (see the CPD plot) and are pulled more closely towards the center of the CGQD (see the EHPD plot). Another prominent feature in Fig. 5 is that $E_B$ in the CGQD can be of the order of 100 meV, which is much larger than that in the CSQD with even smaller radius (about $15\sim50$ meV for a range of dot radii from 2 nm to 15 nm) \cite{Janssens2001}. This large binding energy is mainly caused by the combined factors of geometric confinement, magnetic confinement, and reduced screening.

\begin{figure}[t]
\begin{center}
\includegraphics[width=0.48\textwidth]{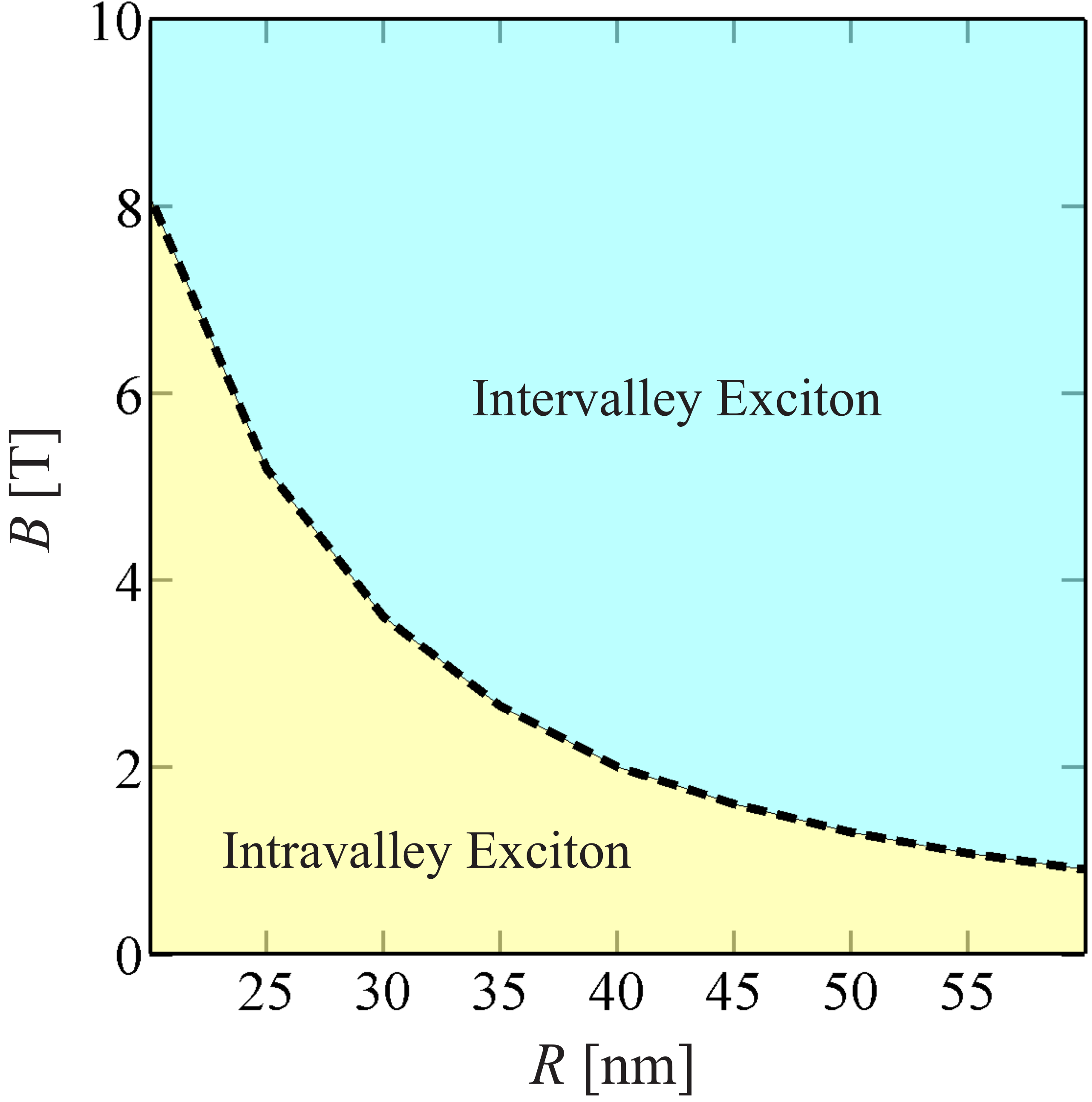}
\caption{$B-R$ phase diagram for the intravalley to intervalley exciton transition in the CGQD. The black dashed curve represents the dependence relation of the critical magnetic field $B_c$ with the dot radius $R$.} \label{Fig7}
\end{center}
\end{figure}

In Fig. 7, we show the $B-R$ phase diagram for the intravalley to intervalley transition of the exciton ground state in the CGQD. Here, the black dashed curve represents the dependence relation of the critical magnetic field $B_c$ with the dot radius $R$. As can be seen, at lower magnetic fields, the exciton ground state is found to be an intravalley exciton state. With increasing field strength to a critical value $B_c$, it changes abruptly into the intervalley exciton state. We find that the critical magnetic field $B_c$ depends on the dot radius $R$ in a manner $B_cR^2\simeq constant$ (see the black dashed curve). This peculiar $B_c-R$ relation can be understood as follows. Since the magneto-exciton states in the CGQD are governed by the two length scales: the magnetic length $l_B=\sqrt{\hbar/eB}$ and the dot radius $R$, the intravalley to intervalley exciton transition occurs when $l_B$ is comparable to $R$, which gives rise to the relation $B_cR^2\sim\hbar/e$. Our numerical result gives $B_cR^2\simeq 3240$ [T$\boldsymbol{\cdot}$nm$^2$].

In Fig. 8, we show the optical transition energies and strengths
for all the exciton states shown in Fig. 3(b). Here, the red circles
denote the results for the intravalley excitons with $T=\pm2$ and $\tau_e=\tau_h=\pm1$, the green circles for the intervalley excitons with $T=0$ and $\tau_e=-\tau_h=1$, and the blue circles also for the
intervalley excitons with $T=0$ but $\tau_e=-\tau_h=-1$. As can be seen, the intravalley and intervalley excitons have different optical transition energies and strengths and exhibit different magnetic field dependencies, as indicated by the color and size of the solid circle in the figure. This magnetic-field tuning of valley-dependent excitonic transitions might shed some light on the potential applications of CGQD in valleytronics. As mentioned previously, intravalley and intervalley excitons are formed in the CGQD due to the valley degree of freedom in graphene [see Fig. 3(b)]. This valley degree of freedom dictates similar optical transitions for the intravalley and intervalley two-electron states in the bilayer graphene quantum dot \cite{Zarenia2013}, where the electron-electron interaction was taken into account.

\begin{figure}[t]
\begin{center}
\includegraphics[width=0.48\textwidth]{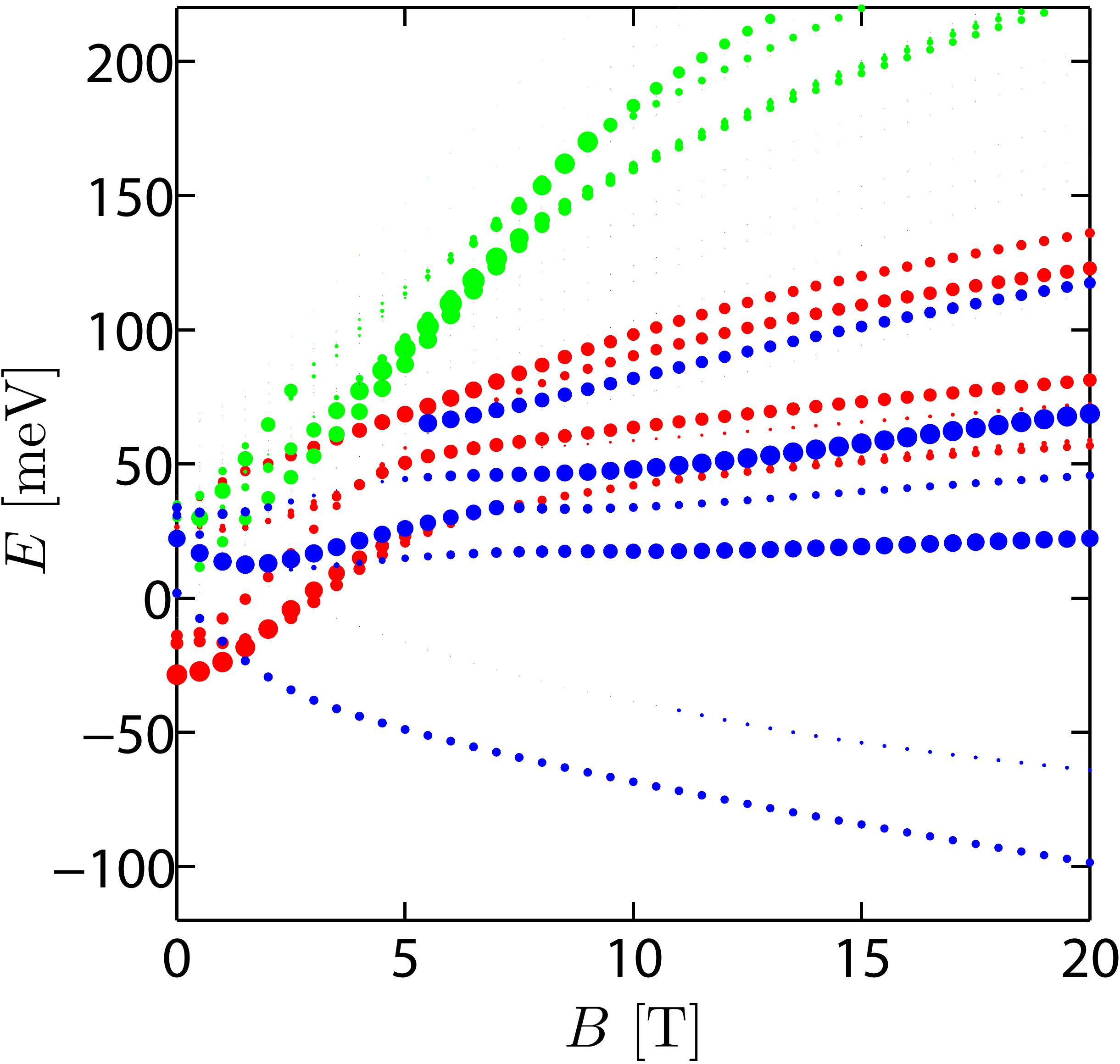}
\caption{Excitonic transition energies and strengths for the same CGQD as in Fig. 2. The red circles denote the results for the intravalley exciton states with $T=\pm2$ and the green/blue circles for the intervalley exciton states with $T=0$. The size of the solid circle indicates the excitonic transition oscillator strength.} \label{Fig8}
\end{center}
\end{figure}

Before closing this paper, we present a qualitative comparison between the optical transitions of exciton states in bulk graphene and in CGQDs. Because bulk graphene has translational invariance, the electron and hole states in the system can be described by 2D plane waves. And due to this invariance, the center-of-mass (COM) momentum for an exciton in bulk graphene is a conserved quantity, which is given by $\hbar\textbf{k}=\hbar\textbf{k}_e-\hbar\textbf{k}_h$ with $\textbf{k}_e$ and $\textbf{k}_h$ being the 2D wave vectors for the electron and hole, respectively. From this expression, we can see that in bulk graphene, the intravalley excitons have zero COM wave vector, i.e., $\textbf{k}=0$, while the intervalley excitons have nonzero COM wave vector, i.e., $\textbf{k}\sim\textbf{K}-\textbf{K}'$, with $\textbf{K}$ ($\textbf{K}'$) being the electron or hole wave vector at the $K$ ($K'$) valley. Because photons have a negligible small momentum, only excitons with $\textbf{k}=0$ are optically active due to the momentum conservation law. Therefore, in bulk graphene, the intravalley (intervalley) excitons are optically bright (dark). However, both the intravalley and intervalley excitons in the CGQD can be optically bright because they have nonzero optical transition strengths (see Fig. 8). Due to the broken translational invariance, the momenta of electrons and holes in the CGQD are no longer good quantum numbers, and so the concept of COM wave vector does not exist for the exciton. Moreover, due to the finite-size effect in QDs, the exciton wave function transforms from the plane-wave form into the envelope-function form. Therefore, new optical transition rules emerge for the intravalley and intervalley excitons in the CGQD, which depend on the overlap between the electron and hole wave functions rather than on the momentum conservation condition in bulk graphene.

\section{Concluding remarks}

We have investigated the exciton states in a CGQD in the presence
of a perpendicular magnetic field. The energy spectrum, wave function, binding energy and oscillator strength of exciton states were calculated within the configuration interaction approach as a function of the magnetic field. We found significant excitonic effects in the CGQD as compared to excitons in the CSQD due to the combined factors of geometric confinement, magnetic confinement, reduced screening and the presence of two valleys. We showed that there are two types of excitons (intravalley and intervalley excitons) in the CGQD because of the valley degree of freedom in graphene, and the intravalley and intervalley exciton states display different magnetic field dependencies due to the different electron-hole symmetries exhibited in the single-particle energy spectra.

With increasing magnetic field, the exciton ground state undergoes an intravalley to intervalley transition accompanied by a change of angular momentum (i.e., a combined transition of valley and angular momentum) and due to this transition, the exciton binding energy changes discontinuously with the magnetic field. Such a combined transition of valley and angular momentum does not occur for the exciton ground state in a CSQD. The exciton binding energy in the CGQD does not increase monotonically with the magnetic field due to the competing geometric and magnetic confinements. We have also examined the optical properties of the exciton states in the CGQD. We found that the optical transition energies and strengths of the intervalley and intravalley excitons can be tuned by the magnetic field. This magnetic-field tuning of the valley-dependent excitonic transitions can be relevant for potential applications of CGQD in valleytronics.

\section{Acknowledgments}
This work was financially supported by the China Scholarship Council
(CSC), the Flemish Science Foundation (FWO-Vl), the National Natural Science Foundation of China (Grant Nos. 11304316, 11574319 and 11604380), and by the Chinese Academy of Sciences (CAS).


\end{document}